\documentclass{amsart}

\usepackage[english]{babel}
\usepackage{amssymb}
\usepackage{amsfonts}
\usepackage{graphicx}
\usepackage[version=4]{mhchem}

\usepackage[margin=1.2in]{geometry}

\newtheorem{theorem}{Theorem}
\newtheorem{corollary}{Corollary}
\newtheorem{lemma}{Lemma}
\newtheorem{example}{Example}
\newtheorem{remark}{Remark}

\DeclareMathOperator{\grad}{grad}
\DeclareMathOperator{\dvrg}{div}
\DeclareMathOperator{\Tr}{Tr}
\DeclareMathOperator{\End}{End}
\DeclareMathOperator{\Hess}{Hess}

\newcommand{\reals}[1]{\mathbb{R}^{#1}}

\newcommand{\tanspace}{\mathrm{T}}
\newcommand{\cotanspace}{\mathrm{T}^{*}}
\newcommand{\volform}{\Omega_{ g }}
\newcommand{\liegroup}[1]{\mathrm{#1}}
\newcommand{\lieder}[1]{\mathcal{L}_{#1}}
\newcommand{\matder}[1]{\frac{\mathrm{d}{#1}}{\mathrm{d}t}}
\newcommand{\parder}[2]{\frac{\partial #1}{\partial #2}}
\newcommand{\LCconn}{\nabla}
\newcommand{\covdiff}{d_{\LCconn}}
\newcommand{\dert}[1]{\frac{d #1}{dt}}
\newcommand{\atPoint}[1]{\big|_{#1}}
\newcommand{\dfrm}{\Delta}
\newcommand{\stress}{\sigma}
\newcommand{\cnj}[1]{{#1}^{\prime}}

\begin{document}
\title{On Dynamics and Thermodynamics of Moving Media}

\author{Anna Duyunova}
\email[E-mail: ]{anna.duyunova@yahoo.com}
\curraddr{V. A. Trapeznikov Institute of Control
Sciences of Russian Academy of Sciences,
65 Profsoyuznaya street, Moscow 117997, Russia}

\author{Valentin Lychagin}
\email[E-mail: ]{valentin.lychagin@uit.no}
\curraddr{V. A. Trapeznikov Institute of Control
Sciences of Russian Academy of Sciences,
65 Profsoyuznaya street, Moscow 117997, Russia}

\author{Serge Tychkov}
\email[E-mail: ]{tychkovsn@ipu.ru}
\curraddr{V. A. Trapeznikov Institute of Control
Sciences of Russian Academy of Sciences,
65 Profsoyuznaya street, Moscow 117997, Russia}



\date{27 November 2024}

\begin{abstract}
In this paper recent results regarding generalized continuum
mechanics on oriented Riemannian manifolds are reviewed and summarized.
The mass, the momentum and the energy conservation laws
are given.
Thermodynamics arising in such media is also considered as a
Lagrangian manifold endowed with a Riemannian structure.
Thermodynamic model of moving media takes into account
deformation and stress arising in a media in motion.
\end{abstract}

\maketitle

\section*{Introduction}
This paper reviews recent research on thermodynamics
and continuum mechanics of moving media possessing some internal
structure.
The principle source of this work is the papers
\cite{duyunova_continuum_2021},
\cite{lychagin_thermodynamics_2022},
\cite{duyunova_2023},
\cite{lychagin_thermodynamics_2023}
and a series of lectures given by the second author.

In Section \ref{sec:Mech1} we discuss fundamental principles
behind the basic equations of fluid dynamics, i.~e.,
the Navier--Stokes and the continuity equations. Thus,
we explain our generalization for the case of Riemannian manifolds.
The equations are given in a coordinate-free form, as well as in
coordinates.
Alternative approaches to generalization of the Navier--Stokes equations
can be found in
\cite{capriz_continua_1989},
\cite{rubin_cosserat_2000},
\cite{vardoulakis_cosserat_2019}.

In Section \ref{sec:therm} we give an overview of
thermodynamics from the point of view of a measurement theory
\cite{lychagin_thermodynamics_2022}. This
leads us to a clear geometrical interpretation of thermodynamic
equations as Lagrangian manifolds. A pseudo-Riemannian structure arising
on these manifolds is considered. We discuss the Gibbs--Duhem
principle that corresponds to the change of information
units in the measurement theory. This principle
allows us to transit to a quotient Legendrian
manifold, which does not include entropy as a coordinate.
Instead, we use a Massieu--Planck potential, which
defines the equation of state. 

In Section \ref{sec:moving} we apply results of the Section
\ref{sec:therm} to a moving medium. In contrast to
thermodynamics of a still medium, our
equations of state includes deformation and stress tensors as
thermodynamic quantities. We consider pseudo-Riemannian structures
arising on such
Lagrangian manifolds. Then, we consider the case
of Newtonian media, which satisfy
the condition that the stress tensor is a linear function
of the deformation tensor.

In Section \ref{sec:eqs}, using the momentum and mass conservation laws we present
an internal energy balance equation, thus we complete
the PDE system for a medium on Riemannian manifolds.
The obtained system together with equations of thermodynamic state
given by Massieu--Planck potential describes motion of such media.


\section{Mechanics}\label{sec:Mech1}
In this section, we discuss basic equations of motion, viz., the continuity
and the Navier--Stokes equations.

Let us start ab ovo, with elementary notions of mechanics.
To describe the motion of a medium, Newton's second law is essentially used. The Navier--Stokes and the Euler
equations are basically this law.

At first, we recall Newton's second law applied
to the motion of an object.
Let $p = m X$ be the momentum of the object, $m$ and $X$ be its mass and
velocity, respectively, and $F$ be a sum of forces
acting on the object. Then Newton's second law reads as
\[
\dert{p}=F.
\]
Certainly, one might understand this equation not as a law of motion,
but as the definition of the momentum $p$ and the force $F$.

The traditional assumption in classical mechanics is that the mass of the object
remains constant during motion, viz., the mass $m$ does not depend on
the velocity $X$, the object is a closed system, and
all phenomena that could change mass, e.~g., radiation, are neglected all together.
Continuum mechanics is used to respect this tradition as well, but in
a more sophisticated form of the {\it continuity equation,}
which will be discussed later.

So, with this assumption at hand, we arrive at the equation
\[
m\dert{X} = F,
\]
which will serve us as the starting point for the Navier--Stokes equation.

The left-hand side of this equation is clear from the geometric point of view. But to
discuss meaningfully the right-hand side $F$, we will require knowledge
of both the thermodynamics and the internal structure of the medium.


Let $(M,  g )$ be an oriented $n$-dimensional Riemannian manifold,
which is considered as the configuration space of a mechanical system, and
$\volform\in\Omega^n(M)$ be the volume $n$-form associated with $ g $.
Note that $ g $ does not exactly correspond to the kinetic energy of the system,
but rather to the specific kinetic energy of the medium.

The medium flow is a time-dependent vector field
$X(t,x)$ on the manifold $M$.

\begin{example}
Molecules of methane \ce{CH4} in a domain $D\in\reals{3}$ can be modeled
as regular tetrahedra in the space. Then, configuration space
$M = D\times(\liegroup{SO(3)}/A_4)$. Here $D$ is the space of mass centers of molecules,
and $\liegroup{SO(3)}/A_4$ is the space of their internal states.
\end{example}

\subsection{Mass conservation law}

Let $\rho (t,x)$ be the mass density of the medium.
Then, $\rho \,\volform$ is considered as the mass
of the infinitesimal volume $\volform$. As in classical mechanics,
we assume that this elementary mass does not change while traveling
in time along the flow, namely,
\[
\left(\parder{\,}{t} + \lieder{ X }\right)\!\left(\rho \,\volform\right)=0,
\]
or
\[
\parder{\rho }{t}\,\volform+X (\rho)\,\volform+
\rho \lieder{X}(\volform)=0,
\]
where $\lieder{ X }$ is the Lie derivative along the field $ X $.

Recalling that, by the definition of a vector field divergence,
$\lieder{X}(\volform) = (\dvrg X)\,\volform$,
we obtain
\[
\left(\parder{\rho }{t}+ X (\rho )+\rho \dvrg X \right)\volform=0,
\]
or, simply,
\begin{equation}\label{eq:cont}
\matder{\rho } + \rho  \dvrg X =0,
\end{equation}
where
\[
\matder{\,}=\parder{\,}{t}+ X 
\]
is a material derivative along the field $ X $.

The equation \eqref{eq:cont} is called the {\it continuity equation.}
As we mentioned earlier, this equation guarantees that the left-hand
side of the Navier--Stokes equation can be
written in the form `$m \times a$'.

One can consider the case when sources or sinks exist within the medium.
Then the continuity equation takes the form
\[
\matder{\rho } + \rho  \dvrg X =S(t,x),
\]
where the `function' $S(t,x)$ characterizes distribution of sources and sinks
in the medium.

\subsection{Acceleration}
Let us consider an infinitesimal element of volume $\volform$,
which moves along the vector field $ X $.

\begin{figure}
\includegraphics[scale=.35]{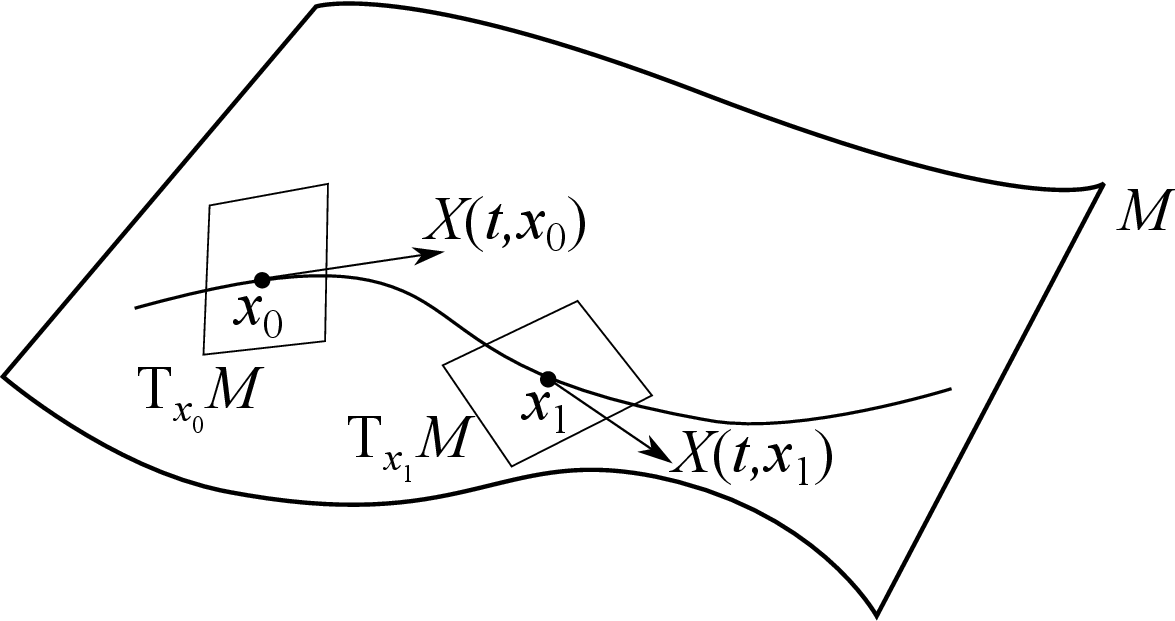}
\caption{To calculate the acceleration we need to compare
vectors from different tangent spaces.}
\label{fig:1}
\end{figure}

Note that to compute the acceleration of this element, one needs not only
the derivative of the velocity with respect to time, i.~e. $\parder{ X }{t}$.
The change of the velocity due to motion of the element along the flow, viz.,
$ X (t,x_1)- X (t,x_0)$, must also be taken into account
(see Figure \ref{fig:1}).

The latter
assumes that there is a way to compare (to subtract) vectors that belong
to the different vector spaces, namely, $\tanspace_{x_1}M$ and $\tanspace_{x_0}M$.
Thus it could be said that we should differentiate the field $ X $
with respect to itself.

To do this, we need a linear connection on $(M, g )$. Note that the Riemannian structure on the configuration manifold $M$, given by the specific kinetic energy $ g $, also supplies us  with a unique torsion-free Levi-Civita connection $\LCconn$, preserving metric $ g $. Thus, as the acceleration $a$ of the elementary volume $\volform$ we take
\[
a=\parder{ X }{t}+\LCconn_{ X } X .
\]

This allows us to write down the left-hand side of Newton's second law as follows
\[
\rho \left(\parder{\,}{t}+\LCconn_{ X }\right)( X )\otimes\volform.
\]
Note that the obtained expression for `$ma$' indicates that the right-hand side, i.~e.
the net force $F$, must be a force applied to the elementary volume. In other
words, a force density is expected there.

\subsection{Coordinate formulation}

Now we write down the above constructions in coordinates.
Let $x=\left(x_1,\dots,x_n\right)$ be local
co\-ordi\-na\-tes on the manifold $M$, and the metric $ g $ be the following
\[
 g  = \sum\limits_{i,j} g _{ij}(x)\, dx_i\otimes dx_j,
\]
where $ g _{ij}= g (\partial_{i},\partial_{j})$ and $\partial_{i}=\parder{\,}{x_i}$.

To describe the connection $\LCconn$, or the directional covariant derivatives
$\LCconn_i=\LCconn_{\partial_{i}}$, $i=1,\dots,n$, we need to define its action
on basis vector fields $\partial_{i}$:
\[
\LCconn_{i}\left(\partial_{j}\right) = \sum\limits_{k}\Gamma^k_{ij}\partial_{k},
\]
where $\Gamma^k_{ij}$ are the Christoffel symbols.

For the Levi-Civita connection, we have
\[
\Gamma^k_{ij}=
\frac{1}{2}\sum\limits_{l} g ^{kl}
\left(
\partial_{i}\left( g _{jl}\right)+
\partial_{j}\left( g _{il}\right)-
\partial_{l}\left( g _{ij}\right)
\right),
\]
where $\| g ^{kl}\|=\| g _{ij}\|^{-1}$ are the components
of the inverse metric.

In what follows, we will write equations in a such manner that
only the Christoffel symbol will be involved but
not the metric $ g $ itself.

The volume form $\volform$ is a unique $n$-form
of  unit length, that is $\left(\volform,\volform\right)=1$, and in coordinates
\[
\volform=\sqrt{\det g}\,dx_1 \wedge\dots\wedge dx_n,
\]
and the divergence of the field $ X $ may also be found with the formula
\[
(\dvrg X)\,\volform=d( X  \mathbin{\lrcorner} \volform).
\]

Though this formula is short, it is not quite useful
for direct computations. We use the Levi-Civita connection
to rewrite it differently and to extend the definition
of divergence to all tensors.

To do this we consider the covariant differential $\covdiff$ corresponding
to the connection $\LCconn$. This differential is a
linear operator
that
\[
\covdiff\colon\mathcal{D}(M) \to
\mathcal{D}(M) \otimes \Omega^1(M),
\]
that acts as follows
\[
\left\langle\covdiff X, Y\right\rangle =
\LCconn_{Y}X,\quad\text{for all}\quad X,Y\in\mathcal{D}(M),
\]
where $\mathcal{D}(M)$ is the $C^{\infty}$-module of
smooth vector fields and $\Omega^k(M)$ is
the module of differential $k$-forms on the manifold $M$.

Note that $\LCconn_Y X$ is $C^{\infty}$-linear in $Y$, and
$\covdiff X$ may be considered as a field of linear operators on
the tangent spaces to $M$. If $X$ stands for the flow
velocity then $\covdiff X$, in some sense, is a linear
approximation of the flow.

Due to presence of the metric $ g $, every linear operator
can be represented as the sum of a self-adjoint and a skew-adjoint
operators. Thus, the self-adjoint part of
$\covdiff X $ is called the {\it rate-of-strain} tensor and
describes deformation (expansion and shear), while the skew-adjoint part
is called the {\it spin} tensor and describes rotation.

Since the spin tensor does not change neither size nor form of
the elementary volume, but only rotates it, we take a closer
look at the self-adjoint part of the operator $\covdiff X $.
This part corresponds to the two phenomena: the expansion and the
shear of the elementary volume. It turns out that the expansion
corresponds to the divergence of $ X $ and is given
by trace of $\covdiff X $ (since traces of the self-adjoint
part and of the operator itself are equal). The action of the expansion part,
simply multiplies the volume by the trace.

Recalling that the divergence of the flow velocity $ X $
shows exactly the same, namely, how the volume of the elementary unit of the medium
changes along the flow,
we get $\dvrg X = \Tr\left( \covdiff X\right)$. The latter can be
proven directly.

In local coordinates $(x_1,\dots,x_n)$ we have
\[
\covdiff X =\sum\limits_{i,j}\left(\partial_{j} X _{i}+
\Gamma^i_{kj} X _k\right)\,\partial_{i}\otimes dx_j
\]
and
\[
\dvrg X = \sum\limits_i\partial_{i} X _i +
\sum\limits_{i,k}\Gamma^i_{ik} X _k.
\]

Therefore, the continuity equation
\eqref{eq:cont} takes the form
\[
\partial_{t}\rho  + \sum\limits_l  X _l\,\partial_{l}\rho  +\rho 
\left(\sum\limits_i \partial_{i} X _i + \sum\limits_{i,k}
\Gamma^i_{ki}  X _k\right) = 0,
\]
or
\[
\partial_{t}\rho +
\sum\limits_l \partial_{l} \left(\rho   X _l\right)+
\rho \sum\limits_{i,k}\Gamma^{i}_{ki} X _k=0.
\]

Respectively, for the acceleration, we have
\begin{align*}
\partial_{t} X  + \LCconn_{ X } X 
=&\sum\limits_l \partial_{t}( X _l)\partial_{l}+\sum\limits_{i,j}
\left( X _i\,\partial_{i}( X _j)\partial_{j}+
\sum\limits_k \Gamma^k_{ij} X _i  X _j \partial_{k} \right) \\
=&\sum\limits_l\left(
\partial_{t} X _l+
\sum\limits_i  X _i\,\partial_{i} X _l+
\sum\limits_{i,j} \Gamma^l_{ij}  X _i  X _j
\right)\!\partial_{l}.
\end{align*}
The first two terms are derivatives of the flow velocity $ X $
with respect to time and along the flow field itself. The third one
is somewhat surprising since it is quadratic in velocity and depends
on Christoffel symbols.

\subsection{Force and divergence}

Finally, we are ready to discuss the right-hand side of Newton's law,
i.~e. the net force acting on the elementary volume $\volform$.
In continuum mechanics, we have two
kinds of forces acting on it. They are volume and surface forces.
A volume force, which can be represented as a product $F(t,x)\,\volform$
of the force density $F$ and the volume form $\volform$,
is quite simple to include into the right-hand side. The primary
example of such a force is gravitation: $\rho \mathbf{g}\,\volform$.

The surface forces such as pressure and internal friction
act rather differently and cannot be included in the same manner.

Thus, the main goal in this subsection is to develop a general framework
for converting the surface forces into the volume ones.

Let $A\in\mathcal{D}(M)\otimes\Omega(M)$ be a field
of a surface force.

To clarify how such forces act, consider
a `small' hyper-surface element $\Delta S$, and let $\mathbf{n}$ be the unit
normal at a point $b\in\Delta S$. Then by a force acting on $\Delta S$
we understand the following vector
\[
A( \Delta S) \overset{\text{def}}{=} A(\mathbf{n}) |\Delta S|\in \tanspace_{b}M,
\]
where $|\Delta S|$ is the `area' of $\Delta S$.

The volume force that corresponds to the surface force is calculated as follows.
We take a small volume $\Delta V$ with the boundary surface $S$. Then $A(dS)$
defines a vector valued $(n-1)$-form on $S$.

In order to find the net force acting on $\Delta V$,
we should calculate an `integral vector sum' of the surface forces
over the boundary $S$: $\int\limits_{S}A(dS)$.
Clearly, this sum is impossible to define, because its summands would
belong to different vector spaces. Again we use the parallel
transport provided by the connection $\LCconn$
to identify vectors in $\tanspace_bM$, $b\in S$,
with vectors in $\tanspace_aM$,
where $a$ is a `central' point of $\Delta V$.

Once collected at the point $a$, forces give us
the density of the volume force corresponding to the surface forces:
\[
\lim_{\Delta V\to 0}\frac{\int\limits_S A(dS)}{\Delta V}.
\]

We denote the density of the volume force
as $\dvrg A$,
corresponding to the field of surface force $A$.
Straightforward computations show that
the divergence is the following operator
\[
\dvrg\colon\mathcal{D}(M) \otimes \Omega^1(M)
\overset{\covdiff}{\to }\mathcal{D}(M) \otimes
\Omega^1(M) \otimes \Omega^1 (M) \overset{c_{1,3}}{\to}\Omega^1(M),
\]
where $c_{1,3}$ is the contraction of the the first and the third
multipliers.

In local coordinates we have
\[
A = \sum\limits_{i,k} A_i^k\partial_{i} \otimes dx_k,\qquad
\dvrg A =\sum\limits_{i,k}
\left( \parder{A_i^k}{x_i}+\sum\limits_j
\left(A_j^k\Gamma_{ij}^i - A_i^j\Gamma_{ik}^j\right)
\right)\!dx_k.
\]
For example, we get
$\dvrg\alpha=d\alpha$ for a scalar operator
$A = \sum\limits_i \alpha(x_1,\dots,x_n) \frac{\partial }{\partial x_i}\otimes dx_i$.

Recall that the left-hand side of Newton's law is a vector field,
and thus we need to transform the differential $1$-form $\dvrg A$ into
the corresponding vector field $\dvrg^{\flat}A$.
To this end, we use the isomorphism given by the metric $ g $, i.~e., $\flat\colon\Omega^1(M) \overset{ g ^{-1}}{\to}\mathcal{D}(M)$.

One may also check that, for $A=X\otimes\omega$, where $X\in \mathcal{D}(M)$ and
$\omega\in\Omega^1(M)$, we have
\[
\dvrg (X\otimes \omega )=(\dvrg X)\,\omega +\LCconn_X\omega.
\]

Summarizing, we see that Newton's law for a continuous medium
has the form of the Navier--Stokes equation:
\[
\rho \left(\partial_{t} X  + \LCconn_{ X } X \right)=
\dvrg^{\flat} A + F,
\]
where $F\in\mathcal{D}(M)$ is a volume force field density, and
$A\in\mathcal{D}(M) \otimes \Omega^1(M)$ is a field of
internal surface forces, existing in the medium.

In general, the force
$A$ depends on the thermodynamics of the medium, which will be the topic of
the next section.

\section{Thermodynamics}\label{sec:therm}
This section is devoted to a geometrical interpretation of the thermodynamics and its role in the measurement theory.

It is known that there are two kinds of quantities
in thermodynamics: {\it extensive} quantities $(E,$ $X_1,$ $\dots,$ $X_n)$ and
{\it intensive} ones $( T , Y_1,\dots,Y_n)$, where $E$ is an internal
energy, and $T$ is a temperature.

\begin{example}
The quantities $(E, V, m)$ are extensive,
and $( T , p ,\eta)$ are the corresponding intensive quantities.
Here $V$, $m$, $p$, $\eta$ are volume, mass, pressure, and
chemical potential, respectively.
\end{example} 

Our goal here is to reformulate the three laws of thermodynamics in pure
geometric terms. But at first, let us recall these laws as one can meet in textbooks.

{\it The first law of thermodynamics,} that is the law of
conservation energy:
\[
\Delta E=\Delta Q - \Delta W,
\]
where $\Delta Q$ is the heat supply to the system,
and $\Delta W$ is a work done by the system.

{\it Second law of thermodynamics:} there exists a function called
entropy $S$ and
\[
\Delta Q =  T  \Delta S.
\]
Note that, to be precise, we should write $\delta Q=T\,dS$,
where $\delta Q$ is a differential 1-form corresponding to the
linear part of $\Delta Q$.

This expression does not clarify what $T$ and $S$ are, it only postulates their
existence. The entropy $S$ may be defined in several ways.

{\it Third law of thermodynamics:} there exists a limit
\[
S_0 = \lim_{T\to 0}S
\]
called the residual entropy.

We outline relation between thermodynamics
and the so-called measurement theory.
This topic was explained in details
in \cite{lychagin_thermodynamics_2022}.

In thermodynamics there exists a distinction between two types of variables:
extensive and intensive variables. The first ones are
averages of a random vector,
where averages are taken with respect to some probability measures.

Variables of the second type are called intensive, and
they `label' extreme probability measures, i.~e., such measures that realize
the principle of minimal information gain (similar to the principle of
maximal entropy).

The other essential part of thermodynamics is the equation of state,
that is, relations between extensive and intensive variables.
We will see that the equations of state are Legendrian
or Lagrangian manifolds. Points on these manifolds are triplets:
an extreme probability measure, an average of a vector variable with respect to this
probability measure, and the information gain.

Thus, essentially, thermodynamics is realized as a trinity
of extensive and intensive quantities, and the
equation of state.

To rewrite the first law in terms of differential forms, we introduce a space $\Phi=\reals{2n+3}$ with
coordinates $(E,X,T,Y,S)$, where $X$ is an $n$-dimensional vector of
the extensive variables and $Y$ is a vector of intensive (dual) variables.
The space $\Phi$ is equipped with the differential form
\[
\omega=dE-T\,dS+Y\,dX.
\]

According to the first law,  we have to 
consider a submanifold $L\subset\Phi$, where the form
$\omega$ vanishes, i.~e. the energy conservation law holds.
Such manifolds that have maximal dimension, $n+1$,
are called Legendrian submanifolds.

This term was suggested by V. Arnold
as an extension of the notion Lagrangian submanifolds
introduced by V. Maslov.

We will call Legendrian manifolds
{\it equations of thermodynamic state.}

To eliminate the entropy we consider
the projection $\pi\colon\Phi=\reals{2n+3}\to\widetilde\Phi=\reals{2n+2}$, where
\[
\pi\colon(E,X,T,Y,S)\mapsto(E,X,T,Y),
\]
and $\widetilde L = \pi(L)$. 

Then, $\widetilde\Phi$ is a symplectic space
with a structure form
\[
\Omega = d(T^{-1}\,\omega) = d(T^{-1})\wedge dE + d(T^{-1}Y)\wedge dX.
\]
Note that we took a form proportional to $\omega$ in order to exclude $S$.

For simplicity,
we assume that $\pi\colon L\to \widetilde L$ is
a diffeomorphism. 
Then $\widetilde L$ is a Lagrangian submanifold in $\widetilde\Phi$, that is,
$\Omega\atPoint{\widetilde L}=0$, and $\dim\widetilde L=\frac 1 2 \dim\widetilde\Phi = n+1$.

Assume now that we have a simply connected
Lagrangian submanifold $\widetilde N\subset\widetilde\Phi$.
Then a differential 1-form
$\theta =  T ^{-1}\,dE+ T ^{-1}Y\,dX$ is closed
on $\widetilde N$ and, therefore, exact. Let $\theta = dF$ on $\widetilde N$. Then the graph of this function, $N = \{S = F\}$, is a Legendrian manidold
in $\Phi$, and $\pi(N)=\widetilde N$.

\subsection{The Maslov lemma and thermodynamics potentials.}

As above, let $L\subset\Phi$ be a Legendrian
manifold and let $\widetilde L\subset(\widetilde\Phi,\Omega)$
be the corresponding Lagrangian
manifold.
For simplicity, we will use
new coordinates in $\Phi$ and $\widetilde\Phi$:
\begin{align*}
&z = S,\quad y_1 = T^{-1},\quad
y_{2} = T^{-1} Y_1,\,\dots,\,
y_{n+1} = T^{-1} Y_n,\,\\
&x_1 = E, \quad
x_{2} =  X_1,\,\dots,\,
x_{n+1} = X_n.\,
\end{align*}

Then the contact 1-form $\theta = -T^{-1}\omega$ and the symplectic form
$\Omega = d\theta$ are
the following
\[
\theta=dz-y\,dx,\qquad \Omega =dx \wedge dy.
\]

By \textit{canonical} coordinates on
$L$ and $\widetilde L$, we mean independent functions $\left(x_{i_1},\dots,x_{i_k},y_{j_1},\dots,y_{j_{n+1-k}}\right)$
such that $\left\{ i_1,\dots,i_k \right\}\bigcap\left\{ j_1,\dots,j_{n+1-k} \right\}=\varnothing$.

\begin{lemma}[V. Maslov]
Any Lagrangian or Legendrian manifold possesses an atlas of canonical coordinates.
\end{lemma}

Now let $\left(x_{i_1},\dots,x_{i_k},y_{j_1},\dots,y_{j_{n+1-k}}\right)$
be global canonical coordinates on the Legendrian manifold $L$. Rewrite
the structure form $\theta$ as follows,
\[
\theta
=dz-\sum\limits_{r=1}^k y_{i_r}\,dx_{i_r}-\sum
\limits_{s=1}^{n+1-k}y_{j_s}\,dx_{j_s}=
d\!\left( z-\sum\limits_{s=1}^{n+1-k}y_{j_s}x_{j_s}\right)-
\sum\limits_{r=1}^k y_{i_r}\,dx_{i_r}+
\sum\limits_{s=1}^{n+1-k}x_{j_s}\,dy_{j_s}.
\]

Then, the condition $\theta\atPoint{L} = 0$ and the condition
that differentials of the functions
\[
\left(x_{i_1},\dots,x_{i_k},y_{j_1},\dots,y_{j_{n+1-k}}\right)
\]
are
independent, give us the following representation of $L$:
\[
z = \varphi +
\sum\limits_{s=1}^{n+1-k}y_{j_s}\parder{\varphi}{y_{j_s}},
\quad
y_{i_r}=\parder{\varphi}{x_{i_r}},
\quad
x_{j_s}=-\parder{\varphi}{y_{j_s}},
\]
where the function $\varphi=\left.\left(
z-\sum\limits_{s=1}^{n+1-k}y_{j_s}x_{j_s}
\right)\right|_L$ is a so-called \textit{free entropy},
or \textit{Massieu--Planck potentials,}
while $y_{i_r}$ and $x_{j_s}$ are called `forces'.

Note that the potential $\varphi$ defines the Lagrangian,
as well as Legendrian,
manifolds completely.

Summarizing, we arrive at the following statement.
\begin{theorem}\label{th:planck}
Let $\left(x_{i_1},\dots,x_{i_k},y_{j_1},\dots,y_{j_{n+1-k}}\right)$
be global canonical coordinates on the Legendrian manifold $L$ and $\varphi$ be the Massieu--Planck potential. Then the Legendrian manifold has the following representation:
\[
z = \varphi +
\sum\limits_{s=1}^{n+1-k}y_{j_s}\parder{\varphi}{y_{j_s}},
\quad
y_{i_r}=\parder{\varphi}{x_{i_r}},
\quad
x_{j_s}=-\parder{\varphi}{y_{j_s}}.
\]
\end{theorem}

\begin{example}\label{exam:3}
In thermodynamics of gases,
the standard
coordinates are $(S, E, V, m, T, p, \eta)$,
$\omega = dE-T\,dS+p\,dV -\eta\,dm$, and
\[
\theta=-T^{-1}\omega = dS - T^{-1}\,dE-T^{-1}p\,dV+
T^{-1}\eta\,dm,
\]
that is,
\[
z=S,\quad
x_1=E,\quad x_2=V,\quad x_3=m,\quad 
y_1=T^{-1},\quad y_2=T^{-1}p,\quad 
y_3=-T^{-1}\eta.
\]

We have the following canonical coordinate
atlases and their corresponding potentials.
\begin{align*}
&x_1, y_2, y_3, \quad \varphi=z - x_2 y_2 - x_3 y_3 =S- \frac{pV}{T} + \frac{m\eta}{T};\\
&x_2, y_1, y_3,\quad \varphi = z-x_1y_1-x_3y_3 = S - \frac{E}{T}+\frac{m\eta}{T};\\
&x_3, y_1, y_2,\quad \varphi = z-x_1y_1-x_2y_2=S - \frac{E}{T} - \frac{pV}{T};\\
&y_1, x_2, x_3,\quad \varphi = z-x_1y_1 = S-\frac{E}{T};\\
&y_2, x_1, x_3,\quad \varphi = z-x_2y_2 = S-\frac{pV}{T};\\
&y_3, x_1, x_2,\quad \varphi = z-x_3y_3 = S +\frac{m\eta}{T};\\
&y_1, y_2, y_3,\quad \varphi = z-x_1y_1-x_2y_2-x_3y_3 = S - \frac{E}{T} - \frac{pV}{T} + \frac{m\eta}{T}.
\end{align*}
\end{example}

\begin{remark}
By the Helmholtz free energy, they mean $H=E-TS$.
Its physical interpretation 
is a work obtainable from a thermodynamic system during an isothermic
process, since
\[
\theta=d(-T^{-1}H)-T^{-2}E\,dT-T^{-1}Y\,dX.
\]
That is, we get the energy conservation law in the form:
$\Delta H=\Delta Q-\Delta E=-\Delta W$,
if $T=const$.

Now consider a Legendrian manifold $L$ with the coordinates $(T,X)$,
then this manifold is defined by relations:
\[
E=h-T\parder{h}{T},\quad Y=-\parder{h}{X},\quad S=-\parder{h}{T},
\]
where $h=H\atPoint{L}$.
\end{remark}

\subsection{Geometrical structures on Lagrangian manifolds}
In this subsection, we consider
Riemannian structures 
arising on Lagrangian manidolds.

As we have seen, equations of state are a Lagrangian manifold
$\widetilde{L}\subset \mathbf{V}^{\ast}\times\mathbf{V}$,
where $\mathbf{V}$ is a vector space of extensive quantities,
and $\mathbf{V}^{\ast}$ is the dual space of intensive quantities.

A point $a\in\widetilde L$ is called {\it regular},
if a projection of the tangent space $\tanspace_a \widetilde L$
to $\mathbf{V}$ is an isomorphism.

Let $a$ be a regular point then $\tanspace_a L$
can be considered as
a graph of a linear isomorphism
$\kappa_a\colon\mathbf{V}\to\mathbf{V}^{\ast}$.

Since $\widetilde L$ is Lagrangian, this isomorphism is a self-adjoint operator.

In other words, the operator $\kappa_a$ defines a non-degenerate
symmetric quadratic form on $\tanspace_a \widetilde L$.

Thus, we get a pseudo-Riemannian structure $\kappa$ on the regular part of $\widetilde L$.

The regular points of $\widetilde{L}$, where $\kappa$ is negative-definite,
we call {\it applicable.}

These points have a transparent
interpretation.
Namely, if we consider thermodynamics as a theory of measurement of extensive quantities \cite{lychagin_thermodynamics_2022},
then $\kappa_a$ coincides with the second central moment.

The set of all applicable points is a union of connected
components, which are called {\it phases.}
Due to Ehrenfest \cite{ehrenfest_phasenumwandlungen_1933}, transitions from one component to another correspond to the
{\it phase transitions of the first order.}

Let us consider a quadratic
differential form $\chi = \sum\limits_{i=1}^{n+1} dx_i\cdot dy_i$. Then, it is easy to check that
$\kappa$ is the restriction of the form
$\chi$ to the manifold $\widetilde L$, and, therefore,
it is defined on the whole manifold $\widetilde L$.
Moreover, singular points are precisely the points,
where the form $\kappa$ is degenerate.

\subsection{Gibbs--Duhem Principle}

The Gibbs--Duhem principle states that Legendrian manifolds
under consideration have to be invariant under the group $Sc$ of scale contact transformations
$\Phi\to\Phi$:
$(z,x_1,\dots,x_{n+1},y_1,\dots,y_{n+1})\mapsto
(tz,tx_1,\dots,tx_{n+1},y_1,\dots,y_{n+1})$, $t>0$.

This group is a group of shifts along the diagonal field
\[
R = z\parder{}{z}+\sum\limits_{k=1}^{n+1} x_k\parder{}{x_k}.
\]

The approach to thermodynamics as a theory of measurement
of extensive quantities shows that this principle
is equivalent to the independence of
equations of states on used units of information \cite{lychagin_thermodynamics_2022}.

It is easy to see that the condition that the Legendrian manifold satisfies the Gibbs--Duhem principle, or invariant with respect to the vector field $R$, is equivalent to the manifold
$L$ belonging to the zeroes of the generating function $F=\theta(R) = z-\sum_{k=1}^{n+1}y_kx_k$.

We call the submanifold $\mathcal{E}=F^{-1}(0)$ as \textit{Euler manifold.}

The quotient $Q = \mathcal{E}/Sc$ is a $(2n+1)$-dimensional
contact manifold. Indeed, the restriciton
$\theta\atPoint{\mathcal{E}}=d(yx) - y\,dx = x\,dy$.

Note that functions
\[
x_k^{\prime} = \frac{x_k}{x_{n+1}}, \quad
z^{\prime} = \frac{z}{x_{n+1}}, \quad (k=1,\,\dots,\,n),\quad
y_1,\dots, y_{n+1} \quad
\]
are integrals of the vector field $R$, and any other integral
is a function of these ones.

Therefore, the functions $x^{\prime}$ and $y$ are 
coordinates on $Q$, and the form $\theta\atPoint{\mathcal{E}}$ is proportional 
to the form
\begin{equation}\label{eq:contact_reduced}
\theta^{\prime} =
dy_{n+1}+\sum_{k=1}^n x_k^{\prime}\,dy_k.
\end{equation}

A quotient manifold $L^{\prime}=L/Sc$ is an $n$-dimensional
submanifold, which is also an integral manifold of the form 
$\theta^{\prime}$ (i.~e., Legendrian).

Also note that the quadratic form $\chi$
is a conformal invariant of the scale group $Sc$,
and the pairing $R \mathbin{\lrcorner}\chi=\sum_{k=1}^{n+1} x_k\,dy_k = x_{n+1}\theta^{\prime}$
vanishes on $L^{\prime}$.

Thus, the restriction $\chi\atPoint{L}$ defines a conformal
structure on submanifold $L^{\prime}$.

Moreover, the quadratic form $\chi = dx\cdot dy$
restricted to the manifold $Q$ 
has the form
\[
\chi =
\sum_{k=1}^n d(x_{n+1}x_k^{\prime})\cdot dy_k + dx_{n+1}\cdot dy_{n+1} =
x_{n+1}\sum_{k=1}^n dx_k^{\prime}\cdot dy_k +
\theta^{\prime}\cdot dx_{n+1}.
\]

Thus, for any Legendrian manifold $L^{\prime}\subset Q$,
we get $\chi\atPoint{L^{\prime}} = x_{n+1}\chi^{\prime}$,
where
\begin{equation}\label{eq:reduced_riemann}
\chi^{\prime} = \sum_{k=1}^n dx_k^{\prime}\cdot dy_k.
\end{equation}

Summarizing, we the contact structure $\theta^{\prime}$
and the pseudo-Riemannian structure $\chi^{\prime}$.

In the case, when $x_{n+1}>0$
on $L$, the applicable domain of $L^{\prime}=L/Sc$ is defined by
the condition $\chi^{\prime }<0$, and the coexistence submanifold is
the border of the domain where $\chi^{\prime }<0$.

\begin{remark}
It is worth to note that if we factor with respect to two
quantities $x_i$ and $x_j$, thus
obtaining contact forms $\theta^{\prime}$ and $\theta^{\prime\prime}$, and quadratic differential forms
$\chi^{\prime}$ and $\chi^{\prime\prime}$,
respectively, then $x_j\theta^{\prime} = x_i\theta^{\prime\prime}$ and
$x_j\chi^{\prime} = x_i\chi^{\prime\prime}$.
\end{remark}

\begin{example}
Let us consider the case of specific quantities of gases.
Quantities describing the thermodynamic state of a gas are $(E,T,V,p,m,\eta,S)$.
To do Gibbs--Duhem reduction, we take $x_{n+1}=m$. Thus, the invariants are
\[
\varepsilon = \frac E m,\quad
v=\frac V m,\quad
\sigma = \frac S m,
\]
which are called specific energy, specific volume, specific entropy, respectively.
We have the Euler equation
\[
\eta m+E-TS+pV=0,
\]
which, essentially, gives us the entropy $S$.

As we have seen in Example \ref{exam:3},
\[
x_1^{\prime}=\frac{E}{m}=\varepsilon,
\quad x_2^{\prime}=\frac{V}{m}=v,\quad x_3=m,\quad 
y_1=T^{-1},\quad y_2=T^{-1}p,\quad 
y_3=-T^{-1}\eta.
\]

Therefore, the structure form of the reduced contact space is
\[
\theta^{\prime} =
dy_3+x_1^{\prime}\,dy_1+ x_2^{\prime}\,dy_2 =
d\!\left(-\frac{\eta}{T}\right) +
\varepsilon\,d\!\left(\frac 1 T\right)
+v\,d\!\left(\frac p T\right),
\]
and
\[
\chi^{\prime} =  dx_1^{\prime}\cdot dy_1
+ dx_2^{\prime}\cdot dy_2 = 
d\varepsilon\cdot d\!\left(\frac 1 T\right)
+dv\cdot d\!\left(\frac p T\right).
\]

Assuming that $T$ and $\rho = v^{-1}$ are global coordinates on the
Legendrian manifold $L^{\prime}$, we get the state
equations in the following form:
\begin{equation}\label{eq:still}
\varepsilon = T^2\psi_{T},\quad
p = -\rho T^2 \psi_{\rho},\quad
\psi = \frac{p}{\rho T} -\frac{\eta}{T}.
\end{equation}
\end{example}

\begin{example}
Let us consider two examples \cite{lychagin_critical_2020}.
Restricting the forms $\chi^{\prime}$
to Legendrian manifolds corresponding to the
ideal gas and the van der Waals gases, we get
the following forms $\kappa$.

    \begin{enumerate}
        \item In the case of ideal gases,
        all points are regular and applicable, and
        the form
        \[
        \kappa = -\frac{Rn}{2T^2}dT^2-\frac{R}{v^2}dv^2
        \]
        is negative-definite.
        \item For the case of van der Waals gases,
        the form has the form
        \[
\kappa = -\frac{Rn}{2T^2}dT^2
-\frac{9R(4Tv^3-9v^2+6v-1)}{4Tv^3(3v-1)^2}dv^2.
        \]
        Thus, applicable points are points, where
        \[
        T > \frac{(3v-1)^2}{4v^3},
        \]
        but $4v^3T=(3v-1)^2$ is a curve
        of singular points, a so-called coexistence curve.
    \end{enumerate}
\end{example}

\section{Thermodynamics of moving media}\label{sec:moving}
By moving media we mean accelerated media, i.~e.,
media that do not move at a constant speed in a straight line.

\subsection{Equations of state}
Thermodynamic quantities describing moving media are divided
into two kinds: scalar and tensor.

For the tensor quantities we use construction similar to ones
we considered in the definition of divergence in Section \ref{sec:Mech1}.

Let $\tau$ be a tensor field of the type $(p, q)$ on
an $n$-dimensional Riemannian
manifold $(M, g)$, and let $\mathcal{O}$ be a `small'
neighborhood of a point $a\in M$, and let $V(\mathcal{O})$ be the volume
of $\mathcal{O}$. With the tensor field $\tau$, we associate a map
$\widehat\tau\colon\mathcal O\to\tanspace^{p,q}_a M$, where
$\widehat\tau(b)\in\tanspace^{p,q}_a M$ is a tensor
obtained from the tensor $\tau(b)$ by parallel transport
along the geodesic connecting the points $a$ and $b$.

By {\it amount} $[\tau]_{\mathcal O}$ of the field $\tau$
contained in $\mathcal O$, we mean
\[
[\tau]_{\mathcal O} = \int_{\mathcal O}\widehat\tau\, \Omega_g,
\]
and, therefore, the density $\rho_a(\tau)$ of the tensor $\tau$ at the point $a$ is defined as
follows
\[
\rho_a(\tau) = \lim_{V(\mathcal{O})\to 0} \frac{[\tau]_{\mathcal O}}{V(\mathcal{O})}
\in\tanspace^{p,q}_a M.
\]

To describe thermodynamics of a medium, we have to specify the
intensive and the extensive quantities, and the Lagrangian manifold
defining the equations of state. 
We assume that the 
equations of thermodynamics state of the medium do not
depend on a point
of $M$. In what follows,
we drop any references to the point $a$ and to the
manifold $M$, namely, we write $\tanspace$ and $\cotanspace$.

Thermodynamics in our model of media is based on
measurement of the following extensive quantities:
mass $m$, volume $V$, internal energy $E$ and
\textit{deformation} tensor $D=\covdiff X$,
where $X$ is the flow velocity field of the medium.

The corresponding dual, or intensive,
quantities are the chemical potential $\eta$,
the pressure $ p $, the temperature $ T $ and
the stress tensor $\stress\in\End \cotanspace$.
In the definition of the stress tensor $\stress$
as a dual to the deformation tensor, we use the duality of $\End \tanspace$
and $\End \cotanspace$ given by the pairing $\langle
A, B\rangle =\Tr A^* B$, where $A\in\End\tanspace$,
$B\in\End\cotanspace$.

As above,
the first law of thermodynamics asserts that the
following differential $1$-form
\[
\theta =
dS
-\frac{1}{T}\,dE
+\frac{1}{T}\langle\stress,dD\rangle
+\frac{\eta}{T}\,dm
-\frac{p}{T}\,dV.
\]

Thus, we have
\begin{align*}
z=S,\quad &x_1=E,\quad x_2 = D,\quad x_3=m,\quad x_4=V,\\
&y_1=\frac{1}{T},\quad y_2=-\frac{\stress}{T},\quad y_3=-\frac{\eta}{T},
\quad y_4 = \frac{p}{T}.
\end{align*}

We apply the Gibbs--Duhem principle to transit to densities of the
extensive quantities, reducing with respect to $x_4=V$ and considering $V\to 0$. Thus, the densities
are $x_1^{\prime} = e$,
$x_2^{\prime} = \dfrm$,
$x_3^{\prime} = \rho$,
which are the energy density $e$,
the rate-of-deformation tensor $\dfrm=\covdiff X\in \End\tanspace$
and the mass density $\rho $.

Thus, contact form \eqref{eq:contact_reduced}
on the quotient manifold
in this case is the following
\[
\theta^{\prime} = d\!\left(\frac{p}{T}\right)
+ e\,d\!\left(\frac 1 T\right)
- \left\langle \dfrm, d\!\left(\frac \stress T\right) \right\rangle
- \rho\,d\!\left(\frac \eta T\right),
\]
and quadratic differential form \eqref{eq:reduced_riemann}:
\[
\chi^{\prime} = de\cdot d\!\left(\frac 1 T\right)
- d\dfrm\cdot d\!\left(\frac \stress T\right) 
- d\rho\cdot d\!\left(\frac \eta T\right)\!.
\]

The Euler equation, in this case, is the following
\[
e-Ts = \Tr( \stress^{*}\dfrm) 
+\eta \rho-p.
\]

Thus, the reduced thermodynamic phase space of the medium is
\[
Q = \reals{5} \times \End \tanspace \times \End \cotanspace,
\]
with coordinates
$(p, e, T, \rho, \eta, \dfrm, \stress)$,
the contact form $\theta^{\prime}$
and the quadratic form $\chi^{\prime}$.

Equations of thermodynamic state of the medium are
Legendrian manifolds $L^{\prime}\subset Q$. 
To describe them we apply the Maslov lemma,
assuming that the functions $T$, $\rho$ and $\dfrm$
are global coordinates on $L^{\prime}$.
Correspondence between coordinates in the Maslov
lemma and the coordinates on $Q$ is as follows
\[
z = \frac{p}{T},\quad x_1= \frac{1}{T},\quad y_1=-e,
\quad x_2= \frac{\stress}{T},\quad y_2=\dfrm,
\quad x_3=\frac{\eta}{T}, \quad y_3= \rho.
\]

From Theorem \ref{th:planck} we get the following statement
(cf. \cite{lychagin_critical_2020}).

\begin{theorem}\label{th:moving}
Any Legendrian manifold $L^{\prime}\subset Q$
with global coordinates $(\rho, T, \dfrm)$
has the following representation
\[
e = T^2\varphi_T,\quad
\stress = -T\varphi_{\dfrm},\quad
\eta = -T\varphi_{\rho}
\]
in terms of the Massieu-Planck potential
\[
\varphi = \frac{1}{T}\left( p -  \Tr( \stress^{*}\dfrm) - \rho\eta\right).
\]
\end{theorem}

This theorem shows that the equation of state is completely defined
by the Massieu--Planck potential $\varphi = \varphi(\rho, T, \dfrm)$.

Then the quadratic form $\chi^{\prime}$
in term of the Massieu--Planck potential is the
following
\[
\chi^{\prime} = -\varphi_{\beta\beta}\,d\beta^2
+ \varphi_{\dfrm\dfrm}\,d\dfrm^2
+2\varphi_{\rho\dfrm}\,d\rho\cdot d\dfrm
+ \varphi_{\rho\rho}\,d\rho^2
= -\varphi_{\beta\beta}\,d\beta^2 + \Hess_{\rho,\dfrm}(\varphi)
< 0,
\]
where $\beta = \dfrac{1}{T}$ is usually called \textit{coldness.}

Note that the relation $\stress = - T\varphi_{\dfrm}$ corresponds to
``Hooke's law'', when $\varphi$ is quadratic in $\dfrm$.
In the next subsection we analyse media such that $\varphi$
is a polynomial in $\dfrm$.

\subsection{Thermodynamic invariants of media}

Assume that a medium possesses a symmetry given by
an algebraic group $G\subset GL(\tanspace)$.

The $G$-action on the tangent space $\tanspace$ can be pro\-long\-ed to a contact $G$-action to the thermodynamic phase space
$Q$, by requiring that this action is trivial on
$\reals{5}=(p, e, T, \rho, \eta)$ and natural
on $\End\cotanspace\times\End\tanspace$.

We consider invariants of this action that are rational functions
of $\dfrm$ with coefficients depending on $(p, e, T, \rho, \eta)$

The Rosenlicht theorem \cite{rosenlicht_remark_1963}
states that there are algebraic invariants $J_1,\dots,J_N$
that generate the field of rational $G$-invariants and separate
regular $G$-orbits. Note that $N$ is equal to the co-dimension
of a regular $G$-orbit.

Then, in the case of $G$-invariant media, i.~e., media possessing
$G$-invariant al\-ge\-bra\-ic Legendrian manifold $L^{\prime}\subset Q$,
we have $\varphi=f(J_1,\dots,J_N)$, where $f$ is a rational function
of $J_1,\dots,J_N$ with coefficients depending on $\rho$ and $T$.

Here we consider in detail so-called Newtonian media, i.~e., media
admitting the symmetry group $G=O( g )\subset GL(\tanspace)$,
where $\tanspace$ is the Euclidean vector space equipped with the metric
$ g $.

The next result is due to Procesi \cite{procesi_lie_2007}.

\begin{theorem}
Algebra of polynomial $O( g )$-invariants is generated by 
invariants
\[
\mathcal{P}_{\alpha,\beta}(A)=
\Tr\,(
A^{\alpha_1}
A^{\prime\beta_1}\cdots
A^{\alpha_m}A^{\prime\beta_m}), \quad
\sum_i(\alpha_i+\beta_i)\leq 2^n-1,
\]
where $A\in\End\tanspace$ is an operator,
$\cnj{A}\in\End\tanspace$ is its adjoint operator with
respect to the metric $ g $, $\alpha=(\alpha_1,\dots,\alpha_m)$,
$\beta=(\beta_1,\dots,\beta_m)$ are multi-indices.
\end{theorem}

We call the invariants $\mathcal{P}_{\alpha,\beta}$
\textit{Artin--Procesi invariants}.

The next result follows from the above theorem and the
Rosenlicht theorem \cite{rosenlicht_remark_1963}.

\begin{corollary}
Field of rational invariants of the $O( g )$-action on
$\End\tanspace$ is generated by any $\frac{n(n+1)}{2}$ algebraically
independent Artin--Procesi invariants. This field separates
regular orbits.
\end{corollary}

Thus, the Massieu--Planck potential of Newtonian media is
a function $\varphi(\rho, T, \mathcal{P}_{\alpha,\beta}(\dfrm))$
rational in $\mathcal{P}_{\alpha,\beta}(\dfrm)$, and equations of state
are the following
\[
e = T^2 \parder{\varphi}{T},\quad
\eta = -T\parder{\varphi}{\rho},\quad
\stress = -T\sum_{\alpha,\beta}\parder{\varphi}{\mathcal{P}_{\alpha,\beta}}
\parder{\mathcal{P}_{\alpha,\beta}}{\dfrm}.
\]

In the case when Newtonian media satisfy ``Hooke's law,'' the
Massieu--Planck potential is a quadratic function:
\[
\varphi=
-\frac{1}{T}\!\left(
\frac{1}{2}\left(
a_{11}(\rho , T )\mathcal{P}_{2}(\dfrm)+
a_{12}(\rho , T )\mathcal{P}_{11}(\dfrm)+
a_{22}(\rho , T )\mathcal{P}_{1}^2(\dfrm)\right)+
b_1(\rho , T )\mathcal{P}_{1}(\dfrm)+
b_2(\rho , T )\right)\!,
\]
where $a_{11}$, $a_{12}$, $a_{22}$, $b_1$, $b_2$ are some functions.

In this case,
the third equation of state takes the form
\[
\stress =
a_{11}(\rho , T )\cnj{\dfrm}+
a_{12}(\rho , T )\dfrm+
(a_{22}(\rho , T )\Tr\dfrm+b_1(\rho , T ))\mathbf{1}.
\]

The next statement follows from Theorem \ref{th:moving}
and equation \eqref{eq:still}.

\begin{theorem}
Assume the state equations are continuous
in $\dfrm$ at $\dfrm =0$, that is, the Lagranginan manifold $L^{\prime}$
for still media coincides with intersection of
the Lagranginan manifold $L^{\prime}$ for moving media and $\dfrm = 0$.

Then, 
$\varphi(\rho, T, \dfrm)\atPoint{\dfrm=0} = \rho\psi(\rho, T)$,
where $\psi$ is the specific Massieu--Planck potential
for a still medium
and $\varphi$ is the  Massieu--Planck potential
density for the same medium in motion.
\end{theorem}

Applying this theorem to a moving Newtonian medium we get
the relation between the function $b_2$ and the
specific Massieu--Planck potential of the still medium 
\[
b_2 = -\rho T\psi(\rho, T).
\]

The function $-b_1(\rho, T)$ is a so-called hydrostatic pressure;
and the functions $a_{ij}$ are various types of viscosity.

Another pressure arising in thermodynamics of still media
with the Massiue--Planck potential $\psi$, in general, differs
from the hydrostatic one.

The commonly used assumption that these pressures are equal leads us
to a relation between the functions $b_1$, $b_2$ and $\psi$:
\[
b_1(\rho, T) = \rho^2 T \parder{\psi}{\rho}, \quad
b_2(\rho, T) = -\rho T\psi(\rho, T).
\]

\begin{remark}
It is worth to note that the number of viscosity
types does not depend on the dimension of the manifold $M$,
and these viscosities depend on $\rho$ and $T$.
\end{remark}

\section{Equations of motion}\label{sec:eqs}

As above, we continue to consider a medium with a configuration space
being an oriented Riemannian manifold $(M, g)$.

Flow of the medium is described by a time-dependendent vector field $X$.

As we have seen, thermodynamics of the moving medium is described by the
following quantities: the mass density $\rho$, temperature $T$,
pressure $p$, chemical potential $\eta$, deformation $\dfrm$,
stress $\stress$ and internal energy density $e$.

The first two equations of motions, we discussed in Section \ref{sec:Mech1}.
The last equation 
expresses the law of energy conservation for an elementary volume.

Let $u$ be the density of total energy of a moving
elementary volume.
Usually (see, for example, \cite{groot_non-equilibrium_1984}),
the law of energy conservation is written as follows
\begin{equation}\label{eq:energygeneral}
\parder{u}{t}=-\dvrg J,
\end{equation}
where $J$ is the total energy flux vector.
This vector is the sum of the convective term
$u X$, the mechanical energy flux
$\stress( X )$ and the heat diffusion term $J_q$,
\begin{equation}\label{eq:energyflux}
J = u X + \stress( X ) + J_q.
\end{equation}

Total energy is a sum of the kinetic
energy and the internal energy of the medium.
Note that the kinetic energy is given by the metric $g$,
thus, we have
\begin{equation}\label{eq:energysum}
u =\rho \frac{g(X, X)}{2} + e.
\end{equation}
Recall that
the momentum conservation equation (Secion \ref{sec:Mech1})
without external force $F$ is the following
\[
\rho\left(\parder{X}{t} + \LCconn_X X\right)= \dvrg^{\flat}\stress ,
\]
and taking inner product with the velocity $X$, we get
the kinetic energy balance equation
\begin{equation}\label{eq:kinetic}
\rho \matder{}\!\left(\frac{g(X,X)}{2}\right)=
\langle \dvrg\stress, X \rangle + g(\stress,\dfrm).
\end{equation}

Combining equations \eqref{eq:energygeneral},
\eqref{eq:energyflux}, \eqref{eq:energysum} and \eqref{eq:kinetic}, 
we get the internal energy balance equation
\[
\matder{e}+e\dvrg X + \dvrg J_q+g(\stress, \dfrm)=0,
\]
Usually, the heat flow vector $J_q$ is given by Fourier's law
$J_q=-\varkappa (\grad T)$,
where $\varkappa\in\End\tanspace$ is
the thermal conductivity of the medium.

Summarizing, we have the following system of PDEs
describing motion of a medium on a Riemannian manifold $(M, g)$:
\[
\left\{
\begin{aligned}
&\matder{\rho}+\rho \dvrg X =0,\\
&\rho\!\left(\parder{X}{t} + \LCconn_X X\right)=\dvrg^{\flat}\stress,\\
&\matder{e}+e\dvrg X + \dvrg J_q+g(\stress, \dfrm)=0,
\end{aligned}
\right.
\]
where the deformation tensor $\dfrm = \covdiff X$,
and a Massieu--Planck potential
of the medium $\varphi$ defines the quantities
\[
\stress= - T \parder{\varphi}{\dfrm},\quad
e = T^2 \parder{\varphi}{ T }.
\]

\textbf{Acknowledgments.}
All three authors are partially supported by RSF Grant no. 21-71-20034.

\end{document}